\RequirePackage{ifpdf}
\ifpdf % We are running pdfTeX in pdf mode
\documentclass[pdftex]{sigma}
\else
\documentclass{sigma}
\fi

\newcommand{\ds}{\displaystyle }

\def\ga{\gamma}

\def\sig{\sigma}

\def\al{\alpha}

\def\eps{\varepsilon}

\def\var{\varphi}

\def\la{\lambda}

\def\Om{\Omega}

\def\be{\beta}
\def\De{\Delta}

\def\l1{{\lambda}_1}

\def\kd{\partial}

\numberwithin{equation}{section}

\begin{document}

\allowdisplaybreaks

\renewcommand{\PaperNumber}{053}

\FirstPageHeading

\ShortArticleName{Lie Symmetries and Criticality of Semilinear
Dif\/ferential Systems}

\ArticleName{Lie Symmetries and Criticality\\ of Semilinear
Dif\/ferential Systems}

\Author{Yuri BOZHKOV~$^\dag$ and Enzo MITIDIERI~$^\ddag$}
\AuthorNameForHeading{Y. Bozhkov and E. Mitidieri}

\Address{$^\dag$~Departamento de Matem\'atica Aplicada,
 Instituto de Matem\'atica,\\
 $\phantom{^\dag}$~Estatistica e
 Computa\c c\~ao Cient\'\i f\/ica, Universidade
Estadual de Campinas - UNICAMP,\\
$\phantom{^\dag}$~C.P. 6065, 13083-970 - Campinas - SP, Brasil}
\EmailD{\href{mailto:bozhkov@ime.unicamp.br}{bozhkov@ime.unicamp.br}}
\URLaddressD{\url{http://www.ime.unicamp.br/~bozhkov/}}

\Address{$^\ddag$~Dipartimento di Matematica e Informatica,
Universit\`a degli Studi di Trieste,\\
$\phantom{^\ddag}$~Via Valerio 12/1, 34127 Trieste, Italia}
\EmailD{\href{mailto:mitidier@units.it}{mitidier@units.it}}
\URLaddressD{\url{http://www.dmi.units.it/~mitidier/}}

\ArticleDates{Received February 01, 2007, in f\/inal form March
20, 2007; Published online March 25, 2007}

\Abstract{We discuss the notion of criticality of
 semilinear dif\/ferential equations and systems, its relations to scaling transformations
 and the Noether approach to Pokhozhaev's identities. For this purpose we propose a
 def\/inition for criticality based on the S. Lie symmetry theory. We show that this
def\/inition is compatible with the well-known notion of critical
exponent by considering various examples. We also review some
related recent papers.}


\Classification{35J50; 35J20; 35J60; 35L70}

\section{Introduction}

 It is well known that the so-called critical exponents are found
 as critical powers for embedding theorems of Sobolev type. They
 can be also viewed as numbers which divide existence and
 nonexistence of solutions for various semilinear dif\/ferential equations and systems involving
 power nonlinearities. Such equations appear when the Frech\'et derivatives of Sobolev
 and $L_p$ norms are considered.

 The aim of this paper is to discuss a notion of criticality of dif\/ferential
 equations, its relations to scaling transformations and the Noether
 approach to
 Pokhozhaev's identities \cite{ye}. Our interpretation is based on the S. Lie symmetry theory of
 dif\/ferential equations \cite{bk, ol,ov,i} and the criticality is considered in terms of
 group invariance. In this sense we propose a
 def\/i\-nition of criticality. Then we show that it is compatible with the notion of critical
 exponent by considering various examples. Although these examples
 are {\it semilinear} dif\/ferential equations and systems, we
 believe that this point of view can trace new directions and
 provide deeper understanding of more general dif\/ferential
 equations and systems.

 To begin with, we recall some results already discussed in \cite{b}. Let us f\/irst consider the following class of
 ordinary dif\/ferential equations for $v=v(r)$, $r>0$:
 \begin{gather}
 \label{e1}  -\big(r^{\al }|v'(r)|^{\be } v'(r)\big)=\mu r^{\ga } |v(r)|^{p-1} v(r) ,
 \end{gather}
 where $\al$, $\be$, $\ga$, $\mu $ and $p$ are real numbers and $v'=\frac{dv}{dr}$. This class was
 introduced in \cite{gb} and studied by Clem\'ent, de Figueiredo and Mitidieri in \cite{cfm}. It contains various
 dif\/ferential equations, e.g. the equations which are radial
 forms of PDE involving the Laplace, $p$-Laplace or $k$-Hessian operators, the Lane--Emden equation,
 Emden--Fowler equation, etc. Such equations come from
 mechanics, astrophysics, general relativity, theories of gravitation, atomic physics and
 quantum mechanics. We shall assume that the parameters satisfy the inequalities
 \[ \mu >0,\qquad \be > -1,\qquad \al -\be -1 >0, \]
 and
 \[ \be +1 < p \leq q^{*} -1,\]
 where
 \[ q^{*}=\frac{(\ga +1)(\be +2)}{\al -\be -1}\]
 is the critical exponent for (\ref{e1}). See~\cite{cfm}.

 For this class it has been observed in \cite{yg1} that a Lie point symmetry of (\ref{e1}) is
 a variational symmetry if and only if $p+1=q^{*}$, the critical exponent. This fact suggests that
 the critical exponent for (\ref{e1}) may as well be def\/ined as the only exponent for which any
 Lie point symmetry of (\ref{e1}) is a variational symmetry. This is a way to def\/ine the critical
 exponent without involving functional analysis. That is, using directly the ordinary dif\/ferential
 equation which occurs in the most cases when reducing the proof of the embedding theorems to
 radially symmetric functions. We shall come back to this point later. We just observe that for the radial form
 \[ {\var }''+\frac{n-1}{r} {\var }' +|{\var }|^{{p-1}}\var =0 \] of the partial
 dif\/ferential equation
 \begin{gather}
 \label{u1} \De u +|u|^{{p-1}}u=0
 \end{gather}
 in ${\mathbb{R} }^n$, $n\geq3$, we have $\al=\ga =n-1$, $\be =0$ and the critical
 exponent is exactly \[ p=\frac{n+2}{n-2}, \]
 the well-known Sobolev exponent. The latter property explains why the Lane--Emden equation
 \[ v''+\frac{2}{r} v' +v^5=0, \]
 for $v=v(r)$, describing a star as a ball of condensed gases, can be explicitly solved (since the critical exponent in
 spatial dimension 3 is exactly 5 and hence any Lie point symmetry is variational reducing the order of
 integration procedure by two). We also recall that the equation~(\ref{u1}) admits positive solutions if and only if $p\geq
 (n+2)/(n-2)$. See~\cite{gs1,gs3}.

 The symmetry approach was applied in \cite{yg2}
 to the radial Lane--Emden system in ${\mathbb{R} }^n$, $n\geq 3$,
 \begin{gather}
  u''+\frac{n-1}{2}u' +|v|^{q-1}v =0,\nonumber \\
 v''+\frac{n-1}{2}v' +|u|^{p-1}u  =0. \label{e2}
 \end{gather}
 The obtained result states that a Lie point symmetry is a variational symmetry of (\ref{e2})
 if and only if the positive numbers $p$ and $q$ are such that the point $(p,q)$ belongs to
 the critical hyperbola
 \begin{gather}
 \label{q1} \frac{1}{p+1} +\frac{1}{q+1} =\frac{n-2}{n}.
 \end{gather}
 By the results of Mitidieri \cite{em0,em,em1} and Serrin--Zou \cite{sz3,sz1,sz2} it follows that
 (\ref{q1}) divides existence and nonexistence of positive
 solutions of (\ref{e2}). Hence the name {\it critical} hyperbola.
 Related existence and nonexistence results for higher order equations and systems
 are contained in  \cite{p2,ps3,rcam} and \cite{em2}. For further details
 concerning Noether symmetries for (\ref{e1}) and (\ref{e2}) see the review
 paper \cite{b} and the references therein. We point out that
 similar ideas and results pertaining to semilinear wave equations
 have appeared in the recent papers \cite{ai} and \cite{al}.

 One can observe that in all these cases the symmetries are {\it
 dilations} of the independent
 and the dependent variables. The scaling transformations of this type play an important role
 since the invariance with respect to such transformations is equivalent to an application of a
 {\it dimensional analysis} argument \cite{bk,ol}.

 In this paper we shall show that similar properties are valid for
 partial dif\/ferential equations and systems which have a
 variational structure. This will allow to def\/ine the
 notion of criticality for such dif\/ferential equations and
 systems. For this purpose we consider the Euler--Lagrange equations
 \begin{gather}
 \label{q2} E(L)=0
 \end{gather}
 for the functional
 \begin{gather}
 \label{q3} J[u] = \int_{\Om } L\big(x, u^{\al }, u^{\al }_{(k)}\big) dx,
 \end{gather}
 where $x\in \Om \subseteq {\mathbb{R}}^n$, $n\geq 1$, $u^{\al }(x)$, $\al =1,2,\dots,m$, are $C^k (\Om )$
 functions, $k\geq 1$, the Function of Lagrange $L= L(x, u^{\al }, u^{\al
 }_{(k)})$ depends on $x, u^{\al }$ and the partial derivatives of $u^{\al }$
up to order
 $k$, and $E=(E_{1 }, \dots,E_m)$ is the Euler operator.

 Now we present the basic def\/inition regarding the criticality of
 the systems $(\ref{q2})$ in the semilinear case.

\begin{definition} Suppose that $(\ref{q2})$ is a
 semilinear system. We say that it is
 critical if there exists a dilation
 \begin{gather}
 \label{q4} X=a_i x_i\frac{\kd }{\kd x_i}+ b^{\al } u^{\al } \frac{\kd }{\kd u^{\al
 }}
 \end{gather}
such that
\begin{gather}
\label{q5} X^{(k)}L +L\sum_{i=0}^{n} a_i=0,
\end{gather}
 where $a_i$, $i=1,\dots,n$, $ b^{\al }$, $\al =1,2,\dots,m$, are real
constants and $X^{(k)}$ is the $k$-th order
 prolongation of $X$.
 \end{definition}

 Above and throughout this paper we assume summation over a
 repeated index: the Latin indices vary from 1 to $n$,
 while the Greek ones -- from 1 to $m$.

 The f\/irst immediate
 observation is that the relation (\ref{q5}) means that $X$ is a
 variational symmetry, that is a symmetry of the action functional
 (\ref{q3}) \cite{bk, ol,ov}. This conclusion follows from the
 inf\/initesimal criterion of invariance \cite{ol,ov}. In this way it
 is clear that the left-hand side of
 (\ref{q5}) is the left-hand side of the Noether identity
 \cite{i, i1} for the particular case of dilations~$X$. The Noether
identity was
 discussed in \cite{ye} and used to obtain
 Pokhozhaev type identities. The choice of critical values in the
 Noether identity allows to obtain the `right' form of the
 Pokhozhaev's identity~\cite{ye} and the corresponding nonexistence results in
 appropriate functional spaces for
 problems which obey certain
 type of homogeneity.

 We note that in the present paper we shall deal only with dilations
  which, as we
 shall show, cover the known cases of critical {\it semilinear} dif\/ferential equations
 and systems. Indeed, all considered examples admit dilations as
 symmetries. In regard to more general dif\/ferential equations
 and systems another possible def\/inition of criticality based on
 the property established and discussed in
 \cite{yg1,yg2,b,yb} relating the variational/divergence symmetries
 of critical dif\/ferential equations and the Sobolev theorem would
 be the following

 \begin{definition} We say that the system (\ref{q2}) is
 critical if any its generalized (Lie--B\"acklund) symmetry is variational or
 divergence symmetry.
 \end{definition}

 However, in order to apply the
 Def\/inition 2 one needs a complete group classif\/ication of the
 considered dif\/ferential equations or systems which for the present
 is not available for some of the examples. The group analysis of
 these cases itself is a subject of another work and applications
 of Def\/inition 2 will be treated elsewhere.

 By a straightforward calculation of the extended inf\/initesimals
 of the dilation $X$ one can see that in more detail the equation
(\ref{q5}) reads
 \begin{gather}
  a_i x_i\frac{\kd L}{\kd x_i}+ b^{\al } u^{\al } \frac{\kd L}{\kd u^{\al
 }} + (b^{\al } -a_i) u_i^{\al } \frac{\kd L}{\kd u_i^{\al }} + \cdots{}\nonumber\\
 \qquad{}+  \left(b^{\al } -\sum_{s=1}^{k}a_{i_s}\right) u_{i_1 i_2 \dots i_k}^{\al } \frac{\kd L}{\kd u_{i_1 i_2 \dots i_k}^{\al }}
 +L\sum_{i=0}^{n} a_i=0.\label{q6}
 \end{gather}
 Although the general solution of the f\/irst order linear partial
 dif\/ferential equation (\ref{q6}) can be easily found, we shall
 proceed in a different way. Namely, we shall consider various
 concrete dif\/ferential equations and systems, and for each of them
 we shall f\/ind the criticality condition in terms of its specif\/ic
 parameters. Typical examples are the following theorems, which are
 among the main new results obtained in the present paper.

 \begin{theorem} Let $F=F(u^1,\dots,u^m)\in C^1({\mathbb{R} }^m)$.
Then the system
 \begin{gather}
  -\De u^{1 } = F_{\ds{u^1}},\nonumber \\
 -\De u^{2 } = F_{\ds{u^2}},\nonumber \\
\cdots\cdots\cdots\cdots\cdots \nonumber\\
  -\De u^{m } = F_{\ds{u^m}} \label{e3}
  \end{gather}
  is critical if and only if
  \begin{gather}
  \label{e4} \sum_{i=1}^{m} u^i F_{\ds{u^i}} = \sigma F,
  \qquad \mbox{where} \quad \sig =\frac{2n}{n-2}.
  \end{gather}
\end{theorem}

 Observe that the equality (\ref{e4}) is the Euler identity for $F$. Hence we have

 \begin{corollary}
 The potential system \eqref{e3} is critical if and only if $F$ is a homogeneous function
  of degree $\sig =\frac{2n}{n-2}$.
  \end{corollary}

 \rm A further result can be stated
 as

 \begin{theorem} Let $H=H(u^1,\dots,u^m ,v^1,\dots,v^m)\in C^1({\mathbb{R} }^m)$. Then
 the system of $2m$ equations
\begin{gather} -\De u^{1 }=H_{v^{1 }}, \nonumber\\
 -\De v^{1 }=H_{u^{1 }},\nonumber\\
 \cdots\cdots\cdots\cdots\cdots \nonumber\\
 -\De u^{m }=H_{v^{m }},\nonumber \\
 -\De v^{m }=H_{u^{m }}\label{q7}
  \end{gather}
  \newpage

 \noindent
 is critical if and only if
 \begin{gather}
 \label{q8} a^{\al } u^{\al } H_{u^{\al }} + (1-a^{\al }) v^{\al } H_{v^{\al
 }} =\theta H,
 \end{gather}
 where $\theta = \frac{n}{n-2}$ and $a^{\al }$, $\al =1,\ldots,m$, are
 real positive constants.\end{theorem}

 \begin{corollary} The system \eqref{q7} is
 critical if and only if $H=H(u^1,\dots,u^m ,v^1,\dots,v^m)$ is a~sum
 of homogeneous functions and identity \eqref{q8} holds.
 \end{corollary}

 In this paper we adopt the following terminology. Let $L$ be a linear
 elliptic dif\/ferential operator in divergence form and $L^{*}$ -- its formally
 adjoint operator. An Euler--Lagrange system of type
 \begin{gather*} L u = F_u(u,v), \qquad
L^{*} v = F_v(u,v)
\end{gather*}
 will be called elliptic potential system, while a system of the
 form
\begin{gather*} L u = H_v(u,v), \qquad
L^{*} v = H_u(u,v)
\end{gather*}
 will be called elliptic Hamiltonian system. (Such terminology has
 been used in analysis, e.g.~\cite{cfm1} and~\cite{cvdv}.) Thus
 the system (\ref{e3}) is an elliptic potential system and the system
 (\ref{q7}) is an elliptic Hamiltonian system.

 Nonexistence results for Hamiltonian systems (\ref{q7})
 were obtained by Mitidieri in \cite{em0,em,em1} using Rellich type
 identities established in the same articles. See also \cite{rcam}.

 This work is a natural continuation of the preceding one \cite{b}
 presented at the 6th International Conference ``Symmetry in
 Nonlinear Mathematical Physics'', June  20--26, 2005, Kyiv, Ukraine.
 Here in Sections 4, 6, 8 we shall illuminate some more or less known results. The results
 obtained in Sections 7, 9--13 are new. We shall also
 review some recent papers \cite{ye,yi1,yi2}. In particular, in
 Section 3, we shall comment on the role of the critical exponents in
establishing of Pokhozhaev's identities
 which will complement the discussion in \cite{ye}. The exposition
 in some parts follows closely the text of the
 original articles. It corresponds to the talk one of us (Y.B.) is going to
 give during the 7th International Conference ``Symmetry in
 Nonlinear Mathematical Physics'',  June 24--30, 2007, Kyiv, Ukraine.

 This paper is organized as follows. In the next section we
 introduce notations and preliminaries. Then in Section 3 we comment on the
 Noether approach to Pokhozhaev identities \cite{ye}. In the subsequent sections
 we apply the basic def\/inition to the following partial dif\/ferential equations and systems:
 nonlinear Poisson equations, $p$-Laplace
 equations, equations involving polyharmonic,
 Baouendi--Grushin and Kohn--Laplace operators,
 elliptic systems of potential, Hamiltonian
 and mixed type, hyperbolic Hamiltonian systems and unbounded Hamiltonian
 systems. For each case we f\/ind the corresponding criticality
 conditions. If the nonlinearities are of power type we show that the
 proposed def\/inition is compatible with the notions of critical
 exponent and critical hyperbola. In this regard we consider some
 model equations and systems. Theorems 1 and 2 are proved in
Sections 9 and 10 respectively.

 \section{Preliminaries}

 In this section we outline very brief\/ly some basic notions and
formulae regarding
 variational properties of dif\/ferential equations and systems as well as Lie groups generators and their extensions.
 For further details, systematic and profound expositions the interested reader is directed
 to \cite{bk,i,ol,ov}.

 We shall suppose that all considered functions, vector f\/ields,
 tensors, functionals, etc.\ are suf\/f\/iciently smooth in order for
 the derivatives we write to exist. The independent variable
 $x\in \Om \subseteq {\mathbb{R} }^n $ -- a bounded or unbounded domain. In this
 work we are mainly interested in group invariance properties of the considered
 dif\/ferential equations and systems. For this reason we shall not
 treat the boundary terms and the regularity of solutions.

 The partial
 derivatives of a smooth function $v=v(x)$ are denoted by subscripts:
 \[ v_i:=\frac{\kd v}{\kd x_i},\qquad  v_{ij}:=\frac{{\kd }^2 v}{\kd x_i\kd x_j}, \]
 etc. We shall also assume summation over a repeated index. The Latin indices vary from 1 to $n$,
 while the Greek ones -- from 1 to $m$. The latter will denote collections of functions,
 e.g. $v^{\al }(x)$.

 We introduce the total derivative operator
\[
D_i =\frac{\kd}{\kd x_i}+u_i^{\al }\;\frac{\kd}{\kd
 u^{\al } } +u_{ij}^{\al } \;\frac{\kd}{\kd u_j^{\al }}+\cdots +u_{i i_1 i_2 \dots i_l}^{\al }\;
 \frac{\kd}{\kd u_{i_1 i_2 \dots i_l}^{\al }} +\cdots,
\]
 where $u^{\al }(x)$ are given functions. (See \cite{bk,ol}.)
  If $v$ is a function of $x$, $u^{\al }$ and the
 derivatives of $u^{\al }$ up to order $k$, then
\[
D_i v=\frac{\kd v}{\kd x_i}+u_i^{\al }\;\frac{\kd v}{\kd
 u^{\al }} +u_{ij}^{\al } \;\frac{\kd v}{\kd u_j^{\al }}+\cdots+u_{i i_1 i_2 \dots i_l}^{\al }\;
 \frac{\kd v}{\kd u_{i_1 i_2 \dots i_k}^{\al }}.
 \]

 The Euler--Lagrange equations, corresponding to the functional
 \[
 J[u] = \int_{\Om } L\big(x, u^{\al }, u^{\al }_{(k)}\big) dx,
 \]
 are given by
 \[
 E_{\al }(L)=\frac{\kd L}{\kd u^{\al }}-D_i\frac{\kd L}{\kd u_i^{\al }} +
 D_i D_j\frac{\kd L}{\kd u_{ij}^{\al } } + \cdots + (-1)^k
 D_{i_1} D_{i_2}\cdots D_{i_k} \frac{\kd L}{\kd u_{i_1 i_2 \dots i_k}^{\al }} =0,
 \]
 where the operator
 \begin{gather}
 E_{\al } = \frac{\kd }{\kd u^{\al }}- D_i\frac{\kd }{\kd u^{\al }_i}
 +D_iD_j \frac{\kd }{\kd u^{\al }_{ij}}+ \cdots +(-1)^{k}D_{i_1}D_{i_2}\cdots D_{i_k}
 \frac{\kd }{\kd u^{\al }_{i_1i_2\dots i_k}} + \cdots
 \end{gather}
 is the $\al $-th component of the Euler operator $E=(E_1,\dots,E_m)$
 which corresponds to the (unknown) function $u^{\al }$. See \cite{bk,ol}.

 Further, consider the dif\/ferential operator
 \[
 X={\xi }^i\frac{\kd }{\kd x_i}+ {\eta }^{\al }  \frac{\kd }{\kd
 u^{\al }}.
 \]
 The functions
 \[
 {\xi }^i = {\xi }^i (x,u) ={\xi }^i (x_1,\dots,x_n, u^1,\dots,u^m)
 \] and
 \[
 {\eta }^{\al }=
 {\eta }^{\al }(x,u)={\eta }^{\al }(x_1,\dots,x_n, u^1,\dots,u^m)
 \]
 are called inf\/initesimals
 of the one-parametric group of point
 transformations generated by $X$, that is the transformation
\begin{gather}
\label{aq1}
 x_j^{*}=x_j^{*}(x, u,\eps ),\qquad u^{*\al } =
u^{*\al }(x, u,\eps ),
\end{gather}
 where $\eps $ is a parameter and
\begin{gather}
\label{aq2} {\xi }^i=\left. \frac{\kd x_{i}^{*}}{\kd \eps } \right
|_{\eps =0},\qquad {\eta }^{\al } = \left. \frac{\kd {u
 }^{*\al }}{\kd \eps } \right |_{\eps =0} .
 \end{gather}
 Given a transformation (\ref{aq1}) one can calculate ${\xi }^i$ and ${\eta }^{\al }$ by
 (\ref{aq2}). And vice-versa, given $x_j$, $u^{\al }$, ${\xi }^i$ and ${\eta }^{\al }$,
 the one-parametric group of point
 transformations (\ref{aq1}) is determined by the unique
 solution of the problem
 \begin{alignat*}{3}
 & \frac{d x_j^{*}}{d \eps }= {\xi }^j (x_i^{*},u^{*\al }),\qquad&&
 \frac{d u^{*\al }}{d \eps }= {\eta }^{\al } (x_i^{*},u^{*\al
 }),&\\
& x_j^{*}|_{\eps =0}=x_j,\qquad && u^{*\al }|_{\eps =0}=u^{\al }.&
 \end{alignat*}
 Henceforth we shall identify the Lie point transformation
 (\ref{aq1}) and its inf\/initesimal generator~$X$.

We associate to $X$ its $k$-th order prolongation $X^{(k)}$ given
by
 \begin{gather}
 \label{g2} X^{(k)}= {\xi }^i\frac{\kd }{\kd x_i}+ \eta  \frac{\kd }{\kd
 u} + {\eta }^{(1)\al }_{i} \frac{\kd }{\kd u_i^{\al }}+\cdots
 + {\eta }^{(k)\al }_{i_1 i_2 \dots i_k} \frac{\kd }{\kd u^{\al }_{i_1 i_2
\dots i_k}},
\end{gather}
 where
 \begin{gather*}
 {\eta }^{(1)\al }_{i}=D_i{\eta }^{\al } -(D_i{\xi }^j)u^{\al }_j, \qquad i=1,2, \dots,n;
\\
{\eta }^{(l)\al }_{i_1 i_2\dots i_l} =D_{i_l} {\eta }^{(l-1)\al
}_{i_1 i_2\dots i_{l-1}}
 - (D_{i_l}{\xi }^{j})u^{\al }_{i_1 i_2
 \dots i_{l-1}j},
 \end{gather*}
with $i_l=1,2,\dots,n$ for $l=2,3,\dots,k$, $k=2,3,\dots$. See
 \cite{bk,ol} for further details. The functions $ {\eta }^{(l)\al }_{i_1 i_2\dots i_l}$ are called extended
 inf\/initesimals.

\begin{definition}  A vector f\/ield $X$ is
 a divergence symmetry of $J[u]$ if there exists a vector function
 $B=(B_1,B_2,\dots ,B_n) $ of $x$, $u$ and its derivatives up to some
 f\/inite order, such that
 \begin{gather} \label{u20}
 X^{(m)}L+LD_i{\xi  }^i=D_iB_i ,
 \end{gather}
 or equivalently,
 \begin{gather}
 \frac{\kd L}{\kd x_i}{\xi }_i + \frac{\kd L}{\kd u}\eta +
 \frac{\kd L}{\kd u_i}(D_i\eta -u_jD_i{\xi }_j) + \cdots\nonumber\\
\label{u19} \qquad{} + \frac{\kd L}{\kd u_{i_1 i_2 \dots
i_m}}[D_{i_1} D_{i_2}\cdots
 D_{i_m}(\eta -u_j{\xi }_j)
 +{\xi }_j u_{ji_1 i_2 \dots i_m } ]+LD_i{\xi }^i=D_iB_i.
 \end{gather}
 If $B=0$, then $X$ is called variational symmetry.
 \end{definition}

 Hence, clearly the relation (\ref{q5}) means that the dilation
 (\ref{q4}) is a variational symmetry.

 \section{On the Noether approach to Pokhozhaev identities}

 The celebrated Pokhozhaev's identity \cite{p1,p2} is an important tool
 in the theory of dif\/ferential equations. Among a big variety of
 applications, it is particularly useful in establishing of
 non\-existence results. Commonly its specif\/ic form for each concrete problem
 is obtained by using ad hoc procedures.

 In \cite{ye} we have recently proposed a general unif\/ied method
 to generate Pokhozhaev identities. This approach is based on the Noether
 identity and the Lie symmetry theory. It has been applied in \cite{ye} to various nonlinear dif\/ferential
 equations and systems choosing transformation parameters assuming critical
 values. The essential points of this method can be summarized as
 follows.

 Let $u^{\al }(x)$, $\al =1,2,\dots,m$, be a set of $C^{2k} (\Om )$
 functions, where $k\geq 1$ and $x\in \Om \subseteq {\mathbb{R} }^n$, $n\geq 1$. We denote
 by $A_k$ the space of
 all locally analytic functions of $x$, $u^{\al }$ and the partial derivatives of
 $u^{\al }$ up to order $k$.
 The elements $f(x,u^{\al },u^{\al }_{(k)})$ of $A_k$ are called dif\/ferential
 functions~\cite{bk,i,ol}.

 Consider a dif\/ferential operator of the form
 \[X={\xi }^i\frac{\kd }{\kd x_i}+ {\eta }^{\al }  \frac{\kd }{\kd u^{\al }} \]
 where
${\xi }^i,{\eta }^{\al } \in A_k$. Let \[L=L(x,u^{\al }, u^{\al
}_{(k)})\in A_k\] be an arbitrary dif\/ferential function. Then
the following identity holds
 \begin{gather} \label{u}
 X^{(k)}L +LD_i{\xi }^i = E_{\al }(L)({\eta }^{\al } - u_j^{\al }{\xi }^j) +
 D_i[ L{\xi }^i + W_i[u, \eta - u_j{\xi }^j]],
 \end{gather}
 where $u=(u^1, \dots,u^m)$, $\eta =({\eta }^1 , \dots,{\eta }^m )$,
 $u_i=(\frac{\kd u^1}{\kd x_i}, \dots, \frac{\kd u^m}{\kd x_i})$, $X^{(k)}$  is the $k$-th order
 prolongation of $X$, $E=(E_{1 },\dots,E_m)$ is the Euler operator and the
 operator $W_i$ is def\/ined in \cite{bk}.

 The identity (\ref{u}) is called the Noether identity
 \cite{i,i1}. It is the corner stone of the approach suggested in~\cite{ye}. The main point of \cite{ye} is the observation that
 the Pokhozhaev's identity for solutions of dif\/ferential equations can be obtained from
 the Noether identity for functions after integration and
 application of the Gauss--Ostrogradskii theorem, with
 account of the boundary conditions.

 It is clear that the crucial step in establishing of the
 Pokhozhaev identities is the choice of the operator $X$ which
 appears in~(\ref{u}). For the semilinear dif\/ferential equations
 and systems considered in~\cite{ye} $X$ was a dilation whose
 parameters assume critical values. In the present work we show
 how to f\/ind such critical parameters. Actually, the use of
 critical values of the equation parameters in obtaining the
 Pokhozhaev's identities is the main motivation to write this
 paper.

 \section{Nonlinear Poisson equations}

 Let $x\in {\mathbb{R} }^n$, $n\geq 3$. It is well known that the
equation
 \begin{gather}
 \label{m01} \De u +f(u)=0
 \end{gather}
 has a variational structure. Its function of Lagrange is given by
 \[
 L=\frac{1}{2} u_j^2- F(u),\qquad F(u)=\int_0^{u} f(z)dz.
 \]

 We shall look for a constant $a$ such that
 \[ X^{(1)}L +n L=0,\]
 where
 \[
 X=x_i \frac{\kd }{\kd x_i} + a u\frac{\kd }{\kd u}
 \]
 and the f\/irst order prolongation of $X$ is given by
 \[
 X^{(1)}= x_i \frac{\kd }{\kd x_i} + a u\frac{\kd }{\kd u} + (a-1) u_i\frac{\kd }{\kd u_i}.
 \]
 By a straightforward
 calculation
 \begin{gather}
 \label{m02} X^{(1)}L +n L =\left ( a +\frac{n-2}{2} \right )u_j^2 - a u f(u)-n F(u).
 \end{gather}
 Let $a=(2-n)/2$. Then by (\ref{m02}) the equation (\ref{m01}) is
 critical if and only if
 \[
 \frac{n-2}{2} u f(u)-n F(u) =0.
 \]
 Hence we have proved

\begin{theorem} The equation \eqref{m01} is
 critical if and only if
 \[
 f(u) = c |u|^{2^{*}-1}u,
  \]
 where $2^{*}= 2n/(n-2)$ and $c$ is an arbitrary constant.\end{theorem}

 For the equation (\ref{m01})
 in a bounded domain $\Om \subset R^n$, $n\geq 3$, with
homogeneous Dirichlet condition $ u=0 $ on $\kd \Om $,
 S.I.~Pokhozhaev \cite{p1} obtained in 1965 the following identity
\[
\int_{\Om } \left[\frac{n-2}{2} uf(u)-nF(u) \right]dx =
-\frac{1}{2}\int_{\kd\Om } |\nabla u|^2(x,\nu )ds,
\]
where $\nu $ is the outward unit normal to $\kd\Om $. This
identity immediately follows from (\ref{m02}) with $a=(2-n)/2$ and
the Noether identity.

 \section[$p$-Laplace equations]{$\boldsymbol{p}$-Laplace equations}

 An argument similar to that presented in the preceding section
 applies to quasilinear equations involving the $p$-Laplace operator ${\De }_p$, $p<n$,
 given by ${\De }_p u := {\rm div}\,(|\nabla u|^{p-2} \nabla u) $.
 The result states:

 \begin{theorem} The equation
\[
 {\De }_p u +|u|^{p^{*}-1}u=0
 \]
 in ${\mathbb{R} }^n$, where $n>p$ and $p^{*}=np/(n-p)$ is the unique critical quasilinear $p$-Laplace
 equation. (The uniqueness is up to multiplying factors of
 $u$.)
 \end{theorem}

 Remark. The above equation should be interpreted in a suitable
 weak form. For identities related to the equation
 \[{\De }_p u + f(u)=0 \]
 where $f: \mathbb{R} \rightarrow \mathbb{R}$ is a given function,
 we refer the interested reader to \cite{gv} where a Pokhozhaev
 identity for $C^{1,\al }$-solutions is obtained. See also~\cite{ps3}.

 \section{Polyharmonic equations}

 In this section we consider the polyharmonic equation
 \begin{gather}
 \label{s6} ({-\De })^k u +f(u)=0
 \end{gather}
 in ${\mathbb{R}}^n$, $n>2k$. It was shown in \cite{yb} that the
 dilation
 \[ Z = x_i \frac{\kd }{\kd x_i} + \frac{2k}{1-p} u\frac{\kd }{\kd u}
 \] is a variational symmetry of
 \begin{gather}
 \label{s4} ({-\De })^k u +|u|^{p-1}u=0
 \end{gather}
 if and only if
 \begin{gather}
 \label{s5} p=\frac{n+2k}{n-2k},
 \end{gather}
 the critical Sobolev exponent. In this case all symmetries of
(\ref{s4}) are divergence symmetries \cite{yb}.

 Let $k$ be an even number. Then, as is well known, (\ref{s6}) is
 the Euler--Lagrange equation of the functional
 \[ \int L = \int \left[\frac{1}{2} |{\Delta }^{k/2}u|^2 - F(u)\right]dx,\qquad
  F(u)=\int_0^{u} f(z)dz.
  \]
 Thus the equation (\ref{q6}) with $m=1$, $u^1=u$, $a^1=a$ assumes
 the following form:
 \begin{gather}
 \label{s7} a u \frac{\kd L}{\kd u} +(a-k)u_{i_1 i_2 \dots i_k} \frac{\kd L}{\kd u_{i_1 i_2 \dots i_k}} +n L=0
 \end{gather}
 since $L$ does not depend on the derivatives of $u$ of order less
 than $k$. Substituting
 \[
 \frac{\kd L}{\kd u_{i_1 i_2 \dots i_k}}=({\De }^{k/2} u) {\delta}_{i_1 i_2} \cdots {\delta }_{i_{k-1} i_k}
 \]
 into (\ref{s7}) we obtain
 \begin{gather}
 \label{s8} \left(a-k +\frac{n}{2}\right)\big({\Delta }^{k/2}u\big)^2 -a u f(u) -n
 F(u)=0.
 \end{gather}
 We choose $a=(2k-n)/2$. Then, by (\ref{s8}), the equation
 (\ref{s6}) is critical if and only if
 \[ \frac{n-2k}{2}u f(u) -n
 F(u)=0.\]
 Hence (\ref{s6}) is critical if and only if
 \[ f(u) = c u^{\frac{n+2k}{n-2k}}, \]
 where $c$ is an arbitrary constant. Thus the following theorem holds:

 \begin{theorem} The equation \eqref{s4} with $p$ given by
\eqref{s5} is the only critical semilinear polyharmonic equation
(since the constant $c$
 can be incorporated into $u$ by the change of the dependent variable
 $u=\mu v$ with $\mu = c^{(2k-n)/(4k)}$). \end{theorem}

 The case $k$-odd can be treated in a similar way.

 For important results related to (\ref{s6}) see \cite{ps3} and
 \cite{ps4}.

 \section[Baouendi-Grushin equations]{Baouendi--Grushin equations}

 Let $x\in {\mathbb{R} }^n$, $y\in {\mathbb{R} }^m$, $n\geq 1,m\geq
1$ and
 $u=u(x,y)\in C^2({\mathbb{R} }^n\times {\mathbb{R}}^m)$ be a scalar function.
 Let $\al >0$ be a real number. Then the generalized
 Baouendi--Grushin operator \cite{bao,gr} is def\/ined by
 \begin{gather}
 \label{g1} {\De }_{L} u ={\De }_{x}u +|x|^{2\al } {\De }_{y} u,
 \end{gather}
 where $ {\De }_{x}u=u_{x_i x_i}$ and $ {\De }_{y}u=u_{y_{\mu } y_{\mu }}$
 are the standard Laplacians in $ {\mathbb{R} }^n$ and ${\mathbb{R} }^m$
 respectively, and $|x|=(x^2_{i })^{1/2}$.

 For recent results and applications of the Baouendi--Grushin operator see \cite{da} and the refe\-rences
 therein. We just recall here that the critical exponent associated to ${\De }_{L}
 $ is $\frac{Q+2}{Q-2} $, where $Q=n+(\al +1) m >2$
 is the so-called homogeneous dimension.

 We consider the following semilinear equation:
 \begin{gather}
 \label{b1} {\De }_{L} u + |u|^{p-1}u = 0.
 \end{gather}
 Formally it is the Euler--Lagrange equation of the functional $J[u]=\int F$ where the
function of Lagrange is given by:
 \[ F = \frac{1}{2} |{\nabla }_x u|^2 +\frac{1}{2}|x|^{2\al }|{\nabla }_y u|^2 -
 \frac{1}{p+1}|u|^{ p+1} = \frac{1}{2} u_{x_i}^2 +\frac{1}{2}|x|^{2\al }u_{y_{\mu }}^2 -
 \frac{1}{p+1}|u|^{ p+1}. \]
 It is easy to see by a straightforward calculation that the
 dilation
 \begin{gather}
 \label{b2}  x^{*}_j=\la x_j, \qquad
y^{*}_{\mu } = {\la }^{\al +1} y_{\mu }, \qquad u^{*}={\la
}^{2/(1-p)} u,
\end{gather}
 is admitted by
the Baouendi--Grushin equation (\ref{b1}), that is, it is a Lie
point symmetry of (\ref{b1}). Then using the inf\/initesimal
criterion of invariance \cite{ol} or performing in the action
integral $J[u]$ the above change of variables, we obtain that the
dilation (\ref{b2}) is a variational symmetry of~(\ref{b1}) if and
only if
\[ p=\frac{Q+2}{Q-2}. \]
Clearly, if $\al =0$ this is the critical Sobolev exponent.

 Now we shall study the criticality of the equation
 \begin{gather}
 \label{ww1} {\De }_{L} u + f(u) = 0.
 \end{gather}
 We aim to clarify for which functions $f$ this equation would be
 critical. For this purpose we consider the dilation
 \[ X=x_i \frac{\kd }{\kd x_i} + (\al +1) y_{\mu } \frac{\kd }{\kd y_{\mu }}
 + a u\frac{\kd }{\kd u}, \]
 where $a$ is a constant to be determined. Then the equation (\ref{q6})
 assumes the following form:
 \[
 \left(a-1 +\frac{Q}{2}\right)|{\nabla }_x u|^2+
 \left(a-1 +\frac{Q}{2}\right)|x|^{2\al }|{\nabla }_y u|^2 -auf(u) - QF(u)=0.
 \]
 We choose $a=(2-Q)/2$. Then (\ref{ww1}) is critical if and only if
 \[
 \frac{Q-2}{2}uf(u)-QF(u)=0.
 \]
 That is, $f(u) =cu^{(Q+2)/(Q-2)}$. The following theorem
 summarizes the above considerations.

 \begin{theorem} Up to some multiplying factors, the equation
 \[
 {\De }_{L} u+ u^{(Q+2)/(Q-2)}=0
  \]
 is the only critical semilinear partial dif\/ferential equation
 involving the Baouendi--Grushin ope\-rator.
  \end{theorem}

 \section[Kohn-Laplace equations]{Kohn--Laplace equations}

 In this section we shall review and comment on some results
obtained in \cite{yi1,yi2}.

As it is well known the Heisenberg group $H^n$ topologically is
the real vector space
 ${\mathbb{R} }^{2n+1}$.
 endowed with the product
 \[(x,y,t)  \big(x^1,y^1,t^1\big) = \left(x+x^1, y+y^1, t+t^1 +2\sum_{i=1}^{n}\big(y_ix^1_i - x_iy^1_i\big)\right), \]
 where $(x,y,t),\big(x^1,y^1,t^1\big)\in {\mathbb{R} }^n\times {\mathbb{R} }^n\times \mathbb{R} =H^n$.
 In the last few decades a signif\/icant number of works treats
 partial dif\/ferential equations on the Heisenberg group $H^n$. In this regard various
 authors have obtained existence
 and nonexistence results for equations involving Kohn--Laplace
 operators. The following equation
 \begin{gather}
 \label{u23} {\Delta }_{H^n} u+f(u)=0,
 \end{gather}
 or equivalently
 \[ u_{x_i x_i}+u_{y_i y_i}+4\big(x^{2}_i+y^{2}_i\big)u_{tt}+4y_i u_{x_i t}-4x_i u_{y_i t}+f(u)=0 \]
 will be called the Kohn--Laplace equation. Here the
 Kohn--Laplace
 operator ${\De }_{H^n}$ is the natural subelliptic Laplacian on $H^n$ def\/ined by
 \[ {\De }_{H^n}=\sum_{i=1}^{n} \big(X^2_i+Y^2_i\big), \]
 where
 \[ X_i=\frac{\kd}{\kd x_i}+2y \frac{\kd}{\kd
 t},\qquad Y_i=\frac{\kd}{\kd y_i}-2x \frac{\kd}{\kd t}.\]
 Recall that in \cite{gl3} Garofalo and Lanconelli established existence, regularity and nonexistence results for the
 Kohn--Laplace equation
 in an open bounded or unbounded subset of $H^n$ with homogeneous Dirichlet boundary
 condition. The existence of weak solutions is proved in \cite{gl3} provided the
 nonlinear term satisf\/ies some growth
 conditions of the form $f(u)= o(|u|^{(Q+2)/(Q-2)})$ as $|u|\rightarrow\infty $,
 where $Q=2n+2$ is the homogeneous dimension of
 $H^n$ (\cite{gl3}). The exponent $(Q+2)/(Q-2)$ is the critical exponent
 for the Stein's Sobolev space \cite{gl3}. The nonexistence results
 follow from remarkable Pokhozhaev identities established in \cite{gl3} for the solutions of
 Kohn--Laplace equations on
 the Heisenberg group. General nonexistence results for solutions of semilinear
dif\/ferential inequalities on the Heisenberg group were obtained
by Pokhozhaev and Veron in~\cite{pv}. In \cite{yi1} a complete
group classif\/ication of Kohn--Laplace equations on $H^1$ is
carried out.

 We observe that the Kohn--Laplace equation is
formally the Euler--Lagrange equation of the functional
 \[ J[u]=\int L,\]
 with
 \begin{gather*}
  L=\frac{1}{2}(X_i u)^2 +\frac{1}{2}(Y_i u)^2 -\int_{0}^{u} f(s)ds\\
\phantom{L}{}= \frac{1}{2} u_{x_i}^2 +\frac{1}{2} u_{y_i}^2 +
2\big(x^2_i+y^2_i\big)u_t^2 +2y_i u_{x_i}u_t-
 2x_i u_{y_i}u_t - \int_{0}^{u} f(s)ds.
 \end{gather*}
Then using the def\/inition of Lie point symmetry of a
dif\/ferential equation, one can show that the scaling
 transformation
\begin{gather*} x_j^{*} =\la x_j, \qquad
 y_j^{*}  = \la y_j, \qquad
 t^{*} ={\la }^2t, \qquad
 u^{*}={\la }^{\frac{2}{1-p}} u \end{gather*}
 is admitted by the equation
 \begin{gather}
 \label{u24} {\Delta }_{H^n} u+ |u|^{p-1}u=0.
 \end{gather}

 Further, following \cite{yi2} and performing this change of variables in the functional $J$, it is easy to see that the dilation
 \[ Z= x_i \frac{\kd}{\kd x_i}+y_i \frac{\kd}{\kd y_i}+2t
 \frac{\kd}{\kd t}+\frac{2}{1-p}u\frac{\kd}{\kd u} \]
 is a variational symmetry if and only if
 \[ p=\frac{n+2}{n}=\frac{Q+2}{Q-2}. \]
 Thus the equation (\ref{u24}) admits the variational symmetry group generated by $Z$ if
 and only if $p$ assumes the critical value. Hence one concludes
 as in Sections 3 and 5 that the following theorem holds.

 \begin{theorem}
 The equation $(\ref{u24})$ with $p=(n+2)/n$ is
 the only critical Kohn--Laplace equation. \end{theorem}

 \section{Elliptic potential systems}

 In this section we prove Theorem~1.

 The function of Lagrange for the potential system (\ref{e3}) is
 given by
 \begin{gather}
 \label{p1} L=\frac{1}{2} u_j^{\al } u_j^{\al
 }-F(u^1,\dots,u^m).
 \end{gather}
 Let $a$ be a constant and consider a dilation of the form
 \[
 X=x_i \frac{\kd }{\kd x_i} + a u^{\al }\frac{\kd }{\kd u^{\al }}.
 \]
 Our aim is to f\/ind out a constant $a$ such that $X$ is a
 variational symmetry of (\ref{e3}), and, hence~(\ref{e3}) would
 be critical (see~Introduction). Substituting (\ref{p1}) into
 the equation (\ref{q6}) we obtain that (\ref{e3}) is critical if
 and only if
 \begin{gather}
 \label{p2} \left(a-1+\frac{n}{2}\right)u_j^{\al } u_j^{\al } - a u^{\al }
 F_{u^{\al }} -n F=0 .
 \end{gather}
 Choosing $a=(2-n)/2$ we conclude from (\ref{p2}) that (\ref{e3})
 is critical if and only if (\ref{e4}) holds.

 \section{Elliptic Hamiltonian systems}

 In this section we prove Theorem 2 and some corollaries.

 The Function of Lagrange for the Hamiltonian system (\ref{q7}) is
 given by
 \begin{gather}\label{x1} L=\frac{1}{2} u_j^{\al } v_j^{\al }-H\big(u^1,\dots,u^m,v^1, \dots,v^m\big).
 \end{gather}

 In order to satisfy the basic def\/inition (see Introduction) we
 shall look for a dilation of type
 \begin{gather}
 \label{x2} X=x_i \frac{\kd }{\kd x_i} + A^{\al } u^{\al }\frac{\kd }{\kd u^{\al }} +
 B^{\al } v^{\al }\frac{\kd }{\kd v^{\al }},
 \end{gather}
 where $A^{\al }$, $B^{\al }$, $\al =1,\dots,m$, are constants to be
 determined later.  By (\ref{q6}) and (\ref{x1}) we obtain:
 \begin{gather}
 \label{x3} X^{(1)} L +nL=(A^{\al }+B^{\al }-2+n)u_j^{\al } v_j^{\al
 }-A^{\al } u^{\al } H_{u^{\al }} -B^{\al } v^{\al } H_{v^{\al
 }}-n H.
 \end{gather}
 Let $ A^{\al }+B^{\al }=2-n$ for $\al =1,\dots,m$. Then by
 (\ref{x3}) and the Def\/inition~1, the Hamiltonian system~(\ref{q7})
 is critical if and only if
 \[ -A^{\al } u^{\al } H_{u^{\al }} -B^{\al } v^{\al } H_{v^{\al }} = n H, \]
 which implies (\ref{q8}) if we denote $a^{\al }= -A^{\al }
 (n-2)$. This completes the proof of Theorem 2.

 Further we consider the particular case $m=1$, $u^1=u$, $v^1=v$,
 $a^{\al }=a$. The condition (\ref{q8}) reads
 \begin{gather}
 \label{x4} a u H_u +(1-a) v H_v = \theta H.
 \end{gather}
 The general solution of this linear f\/irst order partial dif\/ferential equation is
 \[ H=u^{\theta /a} \phi (u^{1-a}v^{-a}),\]
 where $\phi $ is an arbitrary function and $\theta = n/(n-2)$.
 If
 \[H=\frac{1}{q+1} |u|^{q-1}u + \frac{1}{p+1} |v|^{p-1}v \]
 by (\ref{x4}) we have that the corresponding Lane--Emden system
\begin{gather}
\label{x5}  -\De u=|v|^{p-1}v, \qquad -\De v= |u|^{q-1}u ,
\end{gather}
 is critical if and only if
 \[
 a u^{q+1}+ (1-a)v^{p+1} = \frac{n}{n-2}\left (\frac{1}{q+1} u^{q+1} + \frac{1}{p+1}
 v^{p+1}\right ).
 \]
 Hence the following theorem holds:

 \begin{theorem} The system \eqref{x5} is critical if and only if
 \[
 \frac{1}{p+1} +\frac{1}{q+1} =\frac{n-2}{n},
 \]
 that is, if and only if $(p,q)$ belongs to the critical hyperbola \eqref{q1}.
 \end{theorem}

 \section{Mixed systems}

 Analogously to the previous two sections we prove

 \begin{theorem} The mixed Hamiltonian-potential system consisting of $2m+r$
equations
 \begin{gather}  -\De u^{1}=H_{v^{1}},  \nonumber\\
 -\De v^{1 }=H_{u^{1}}, \nonumber\\
 \cdots\cdots\cdots\cdots\cdots \nonumber\\
 -\De u^{m }=H_{v^{m}}, \nonumber\\
 -\De v^{m }=H_{u^{m}}, \nonumber\\
 -\De w^{1 } =  H_{w^{ 1}},\nonumber \\
 \cdots\cdots\cdots\cdots\cdots \nonumber\\
 -\De w^{r } =  H_{w^{r }},\label{m13}
 \end{gather}
where $H=H(u^1,\dots,u^m,v^1,\dots,v^m, w^1,\dots,w^r)$,
$H(0,\dots,0)=0$, is critical if and only if
\[ a^{\al } u^{\al } H_{u^{\al }} +(1-a^{\al }) v^{\al } H_{v^{\al }}+
\frac{1}{2} w^{\al } H_{w^{\al }}=\frac{n}{n-2} H, \] where $\al
=1,\dots,m$ and $\be =1,\dots,r$.\end{theorem}

 \section{Hyperbolic Hamiltonian systems}

 \begin{theorem} The nonlinear hyperbolic system of Hamiltonian
type
\begin{gather*} u_{tt}-\De u + H_v(u,v) = 0, \\
v_{tt}-\De v + H_u(u,v)=0 \end{gather*}
 is critical if
and only if
\[ a u H_u+(1-a)v H_v =\frac{n+1}{n-1} H. \]
\end{theorem}

\rm This result is obtained by the same argument as before and we
omit the corresponding details pointing out that in the particular
case of power nonlinearity
\[H=\frac{1}{q+1} |u|^{q-1}u + \frac{1}{p+1} |v|^{p-1}v \]
the latter condition reads
\[ \frac{1}{p+1} +\frac{1}{q+1} =\frac{n-1}{n+1}. \]

 The Pokhozhaev's identity corresponding to the hyperbolic Hamiltonian
 system in Theorem~10 was obtained in \cite{ye} using the
 Noetherian approach. For a discussion on specif\/ic numbers
 concerning the scalar case see~\cite{ai,al}.

 \section{Unbounded Hamiltonian systems}

 \begin{theorem} The system
 \begin{gather*} u_t-\De u = H_v(u,v), \\
-v_t-\De v = H_u(u,v) \end{gather*} is critical if and only if
\[ a u H_u+(1-a)v H_v =\frac{n+2}{n} H. \]
\end{theorem}

\rm Again this result as well as the corresponding Pokhozhaev's
identity \cite{ye} is obtained by the same arguments as before and
we omit further details.  We observe that in the particular case
of power nonlinearity
\[H=\frac{1}{q+1} |u|^{q-1}u + \frac{1}{p+1} |v|^{p-1}v \]
the criticality condition reads
\[ \frac{1}{p+1} +\frac{1}{q+1} =\frac{n}{n+2}. \]
The latter condition appears in \cite{cfm1}, see also \cite{cvdv}.

 \subsection*{Acknowledgements}

 We wish to thank the referees for their useful suggestions. Yuri
 Bozhkov is grateful to the
 Organizers of the 7th International
 Conference ``Symmetry in Nonlinear Mathematical Physics'',  June  24--30
2007, Kyiv, Ukraine, for having given him the opportunity to
 present a
 talk on this subject. He
 would also like to thank FAPESP, CNPq and FAEPEX-UNICAMP, Brasil,
 as well as ICTP, Trieste, Italy, for f\/inancial support. Enzo
 Mitidieri acknowledges the support of INTAS-05-100000B-792.

\pdfbookmark[1]{References}{ref}
\LastPageEnding
 \end{document}